# Selectivity filter gate versus voltage-sensitive gate: A study of quantum probabilities in the Hodgkin-Huxley equation

N. Moradi,[1] F. Scholkmann,[2] V. Salari[1]

[1] Department of Physics, Isfahan University of Technology, Isfahan, Iran
[2] University Hospital Zurich, Biomedical Optics Research Laboratory, Division of Neonatology, 8091 Zurich, Switzerland

vahidsalari@cc.iut.ac.ir

**Abstract**

The Hodgkin-Huxley (HH) model is a powerful model to explain different aspects of spike generation in excitable cells. However, the HH model was proposed in 1952 when the real structure of the ion channel was unknown. It is now common knowledge that in many ion-channel proteins the flow of ions through the pore is governed by a gate, comprising a so-called "selectivity filter" inside the ion channel, which can be controlled by electrical interactions. The selectivity filter is believed to be responsible for the selection and fast conduction of particular ions across the membrane of an excitable cell. Other (generally larger) parts of the molecule such as the pore-domain gate control the access of ions to the channel protein. In fact, two types of gates are considered here for ion channels: the "external gate", which is the voltage sensitive gate, and the "internal gate" which is the selectivity filter gate (SFG). Some quantum effects are to expected in the SFG due to its small dimensions, which may play an important role in the operation of an ion channel. Here, we examine parameters in a generalized model of HH to see whether any parameter affects the spike generation. Our results indicate that the previously suggested semi-quantum-classical equation proposed by Bernroider and Summhammer (BS) agrees strongly with the HH equation under different conditions and may even provide a better explanation in some cases. We conclude that the BS model can refine the classical HH model substantially.

## 1. INTRODUCTION

MacKinnon was awarded the Nobel Prize in 2003 for obtaining the real structure of the potassium crystallographically-sited activation (KcsA) ion channel. The 3.4 nm long KcsA channel comprises a 1.2 nm long selectivity filter that is composed of four P-loop monomers and whose structure is similar to that of the alpha-helix. The difference is due to the fact that they do not have hydrogen bonds that connect conforms (i.e. locally rigid regions) to each other in an alpha-helix structure. Each P-loop is composed of five amino acids: [T (Threonine, Thr75), V (Valine, Val76), G (Glycine, Gly77), Y (Tyrosine, Tyr78), G (Glycine, Gly79)] linked by peptide units (H-N-C=O) in which N-C=O is an amide group and C=O is a carbonyl group. Carbonyls are responsible for trapping and displacement of the ions in the filter. In this paper, we would like to consider the role of the selectivity filter in the Hodgkin-Huxley equation.

## 2. WHY SHOULD WE CONSIDER THE SELECTIVITY FILTER AS A GATE?

There are several reasons why the selectivity filter (SF) should be considered as a gate in the ion channel. First, permeation and gating mechanisms are coupled; the SF that controls permeation is also responsible for opening and closing the channel (Chapman et al., 1997). In the closed states the ions are trapped but in the open states the bound ions are released. The KcsA structure with two K ions in the SF represents the closed conformation (Clapham, 1999). The SF has a different conformation in the open and closed state. In the open state, single KcsA channels are poorly ion selective, permeate partially hydrated K ions, have a wider diameter than seen in the crystal structure (Meuser et al., 1999). Moreover, permeant ions bind with high affinity in the pore. This was first described for $Ca^{2+}$ ions in Ca channels (Armstrong & Neyton, 1991; Polo-Parada & Korn, 1997). Furthermore K ions also bind with high affinity in the K channel pore: micromole K concentrations block Na conductance (Kiss et al., 1998; David et al., 2000). Correspondingly, short closed times in single channel records result from K ions acting as pore blockers (Choe et al., 1998). Additionally, an alternative is needed for the cytoplasmic constriction acting as a gate, since it is not universally found. Inward rectifying K channels have a wide internal entrance (Lu et al., 1999). Glutamate receptors, which have an inverted topology, have a wide external vestibule (Kuner et al., 1996). There is a strong coupling between sensor movement and the conformation of the SF. The effect of mutations in S4 on activation properties depends critically on whether the SF contains a Val or Leu at position 76. Open state stability is determined by the permeating ion species, linking gating to selectivity (Spruce et al., 1989). Also, mutations in the SF affect single channel gating. In the NMDA receptor, a conserved Asparagine residue critical for Ca permeability and Mg block, stabilizes subconductance levels (Schneggenburger & Ascher, 1997). The direction of the K flux determines the open state stability in drk1 and which (sub) conductance levels predominate in KcsA (Meuser et al., 1999). Open state stability depends on the direction of K flux. And finally, the SF makes a better gate because of energy considerations; single channel gating is highly reversible and the closed-open transition requires little free energy.

In general, we have two types of gates for ion channels: an external gate which is the voltage sensitive gate (VSG), and an internal gate, which is the selectivity filter gate (SFG). The flow of ions through the pore is governed by SFG inside the ion channel. Here, we focus on the role of SFG in the channel and replace the VSG with SFG in the Hodgkin-Huxley (HH) Equation (Hodgkin & Huxley, 1952) to examine the results.

## 3. QUANTUM EFFECTS IN THE SELECTIVITY FILTER GATE

Recently, Vaziri & Plenio (2011) and Ganim et al. (2011) proposed the presence of quantum coherence by arguing that the backbone structure of the SF is not rigid as expected in classical models and studied vibrational excitations in K+ ion-channels. They discussed a possible emergence of resonances at the picoseconds (ps) scale in the backbone amide groups that can play a role in mediating ion-conduction and ion-selectivity in the SF. Summhammer et al. (2012) have shown that a quantum mechanical calculation is needed to explain a fundamental biological property such as ion-selectivity in trans-membrane ion-channels. If quantum effects do play a critical role in filter-ion coordination, it is feasible that these delicate interactions could leave their quantum traces in the overall conformation and the molecular gating state of the entire protein (Summhammer et al., 2012). Bernroider and Summhammer (BS) (2012) considered the role of SFG in the HH equation (Hodgkin & Huxley, 1952) and have concluded that there are significant quantum effects in the filter which can affect the initiation of the spike generation. The main motivation for a quantum mechanical approach to ion conduction mechanisms stems from the small dimensions and short range forces among and between ions and the surrounding atomic structure of the ion hosting protein (Bernroider & Summhammer, 2012; Salari et al., 2011). Here, we would like to investigate such a quantum model of the HH equation in more detail.

## 4. Gate variables: classical or quantum mechanical probabilities?

In the HH equation (Hodgkin & Huxley, 1952) the gating variables m, n and h have values between 0 and 1 and represent the probability that the specific condition for the conduction of ions through a membrane channel are met at a given time and transmembrane voltage. For example, the variable m is time dependent and satisfies in

$$\frac{dm}{dt} = \alpha_m(V)(1-m) - \beta_m(V)m.$$

In summary, m indicates the probability that the condition is true and (1-m) is the probability that the condition is false, which depends on the voltage. In general, the classical probability (CP), m, is expressed in terms of voltage and depends on a VSG-based operation. According to the BS model (Bernroider & Summhammer, 2012), entanglement exists between ions, so the variable m can be expressed based on the presence of ions in the SFG. The question is that whether there is any relation between the quantum probabilities (QPs) and the voltage of the channel. Ions outside a SF move in water with an electric permeability $\varepsilon_{water} = 80$, but in a SF the electric permeability is significantly lower (Chung et al., 1999). This leads to an energy barrier that is known as the Born energy:

$$E_B = \frac{q^2}{8\pi\varepsilon_0 R_B}(\frac{1}{\varepsilon} - \frac{1}{80}).$$

For example, for potassium ion with $R_B = 1.93 A^0$ and with $\varepsilon = 20$, $\varepsilon = 40$ and $\varepsilon = 60$, the Born energy will be *5.4 kT, 1.8 kT* and *0.6 kT*, respectively (Chung et al., 1999). Furthermore, Brownian dynamics simulations have shown that at T = 300K when there one, two or three ions in a SF the confining energy scales are between 10 kT (for three ions) and 60 kT (for one ion) (Berneche & Ruox, 2001; Chung et al., 2002). This indicates that in the absence of any electrostatic forces ions are not able to traverse the channel via their thermal energy KT. Also, it may be possible for the quantum tunneling of the ions through the potential of the SFG. If the opening and closing of the SFG depend on the tunneling of the ions then it should be found its relation to voltage, as in the case in the resonant- tunneling diodes where the quantum tunneling probabilities of electrons depend on the voltage (Ionescu & Reil, 2011).

## 5. METHOD

The size of the membrane patch was assumed to be 60μm× 60μm. The solutions of the HH and the modified equations of HH are obtained and plotted by MAPLE software. The assumptions for solving the equations are: C = 1μF/cm$^2$, $E_K$ = 50mV, $E_{Na}$ =−77mV, $E_L$ = −54mV, $g_K$ = 36mS/cm$^2$, $g_{Na}$ = 120mS/cm$^2$, $g_L$ = 0.3mS/cm$^2$, and (in units of 1/ms with voltages in mV), $α_n(V) = 0.01(V + 55)/(1 − \exp(-V - 55)/10)$, $β_n(V) = 0.125\exp(-V - 65)/80$, $α_m(V) = 0.1(V + 40)/(1 − \exp(-V - 40)/10)$, $β_m(V) = 4\exp(-V - 65)/18$, $α_h(V) = 0.07\exp(-V - 65)/20$ and $β_h(V) = 1/(1 + \exp(-V - 35)/10)$, where $g_K$, $g_{Na}$ and $g_L$ are the conductance through the membrane for $K^+$, $Na^+$ and leakages of all other ions, respectively. $E_K$, $E_{Na}$ and $E_L$ are the corresponding equilibrium potentials. C gives the capacitance of the patch of membrane, and the membrane potential by V. $I_{ext}$ denotes an external current, which can arise from a pulse from a neighboring piece of membrane. An external current pulse of trapezoidal shape triggered the pulse.

## 6. What mathematical term can specify the ion configuration?

It is more important to correct sodium channels than the potassium channel, since voltage-dependent sodium channels control the onset dynamics of a HH type action potential. Also, the structure of sodium channels is not well resolved yet; its structure is considered the same as KcsA.

In general, there are five sites in the SFG as (S0 S1 S2 S3 S4) in which water molecules (W) and sodium ions (Na) can be embedded in the SFG by specific configurations. We note here that ions cannot be next to each other in two adjacent sites since the repulsion force causes at least one site distance between them. There is always a water molecule between the ions (Zhou & Mackinnon, 2003). For example, when we have three sodium (Na) ions in the SFG we have only one configuration (NaWNaWNa), which is the correct state for a current in the channel. In this state the channel is open.

## 7. THe Bernroider-Summhammer (BS) model

Bernroider and Summhammer (2012) have considered sodium channels and interpreted the classical probability (CP), m, as a quantum probability (QP) in which there is a possibility that filter states could play a role. The BS model claims that the all possible states can be written in a coherent superposition state

$$|\psi\rangle = \sqrt{(1-m)^3}|0\rangle + \sqrt{3m(1-m)^2}|1\rangle + \sqrt{3m^2(1-m)}|2\rangle + \sqrt{m^3}|3\rangle \quad (1)$$

where (1 -m) is the probability of not being a correct state and m is the probability of being a correct state in the SF. The term $m^3$ means that the channel is open and we have current. For other cases there is no current and the SFG is closed. They have obtained the quantum mechanical correction for the original HH-equation by replacing the term $m^3$ with $\delta^2 m^3$. The effective value of $\delta^2$ is likely to vary stochastically within its range during a neuronal pulse. They replace the parameter $\delta^2$ by another parameter k, which expresses the degree of entanglement, which takes the values between -1 and 1. The case k = 0 represents the classical situation of $\delta^2 = 1$, with no entanglement. The case k = 1 represents maximum positive entanglement, with $\delta^2 = 1 + \frac{1-2m^2+m^3}{m^3}$. Consequently, the BS equation is in the form of Eq. (2),

$$C\frac{dV}{dt} = I_{ext} - g_K n^4(V-E_K) - g_{Na}\left[1+k\frac{3m(1-m)}{1-3m(1-m)}\right]m^3 h(V-E_{Na}) - g_L(V-E_L). \quad (2)$$

## 8. The modified Bernroider-Summhammer (MBS) model

The BS model takes the number of degrees of freedom of the open condition as basis vectors. Here, we would like to consider another form of the equation based on QPs in terms of ions configuration, to see whether changing the configuration changes the results. Here, we assume that the probability of existence of one ion in the SFG is m, while the probability of there being no ion is 1 - m.

When we have three ions in the SFG, i.e. $|3\rangle$, there is only one possibility (NaWNaWNa) meaning the open state of the gate and we have current, so the probability is $(m).(m).(m) = m_3$. When we have two ions in the SFG, i.e. $|2\rangle$, we have only two probable configurations (WNaWNaW), (NaWNaWW) (Zhou & Mackinnon, 2003), so the probability is $2(m).(m).(1-m) = 2m^2(1-m)$. The probability of there being one ion, $|1\rangle$ or no ion, $|0\rangle$, in the filter is very low. Therefore, the total state in a superposition will be in the form of wavefunction (3):

$$|\psi\rangle = \varepsilon|0\rangle + \varepsilon|1\rangle + \sqrt{2m^2(1-m)}|2\rangle + \delta\sqrt{m^3}|3\rangle \qquad (3)$$

where $\varepsilon$ indicates very low probability amplitude. Following the BS-like approach we consider the deviation $\delta$ for the above state

$$|\psi\rangle = \varepsilon|0\rangle + \varepsilon|1\rangle + \sqrt{2m^2(1-m)}|2\rangle + \delta\sqrt{m^3}|3\rangle \qquad (4)$$

By normalization we have $2|\varepsilon|^2 + 2m^2(1-m) + \delta^2 m^3 = 1$. We know that $2|\varepsilon|^2 \geq 0$ and $2|\varepsilon|^2 \leq 1$, hence $1 - 2m^2(1-m) - \delta^2 m^3 \geq 0$, so we can conclude $0 \leq \delta^2 \leq 1 + \frac{1 - 2m^2 + m^3}{m^3}$.

Now, similarly to the BS approach we obtain another equation for positive entanglement as follows:

$$C\frac{dV}{dt} = I_{ext} - g_K n^4(V - E_K) - g_{Na}\left[1 + k\frac{1 - 2m^2 + m^3}{m^3}\right]m^3 h(V - E_{Na}) - g_L(V - E_L). \qquad (5)$$

It is seen that the coefficient $1 + k\frac{3m(1-m)}{1 - 3m(1-m)}$ in the BS model is replaced by the coefficient $1 + k\frac{1 - 2m^2 + m^3}{m^3}$. We call the equation (5) as the modified BS (or MBS) equation. Now, we analyze the equations.

## 9. RESULTS

Here, we would like to compare the above models (HH, BS and MBS) for different parameters.

### 9-1. Gate variable diagrams

The gate variables m, n and h are plotted for the HH, BS and MBS models in the Figure 3. It is seen that for the case k = 0.01 we have repeating rings for MBS, but there is a rather high overlap between BS and HH models. For other values of k, again MBS model has deviations from HH and BS models.

### 9-2. Voltage diagrams

The voltage diagrams versus time are plotted in the Figure 4. Again, MBS model is obviously different while there is a high overlap between BS and HH models. It is seen that for the k values bigger than 0.1 the number of pulses are the same as each other in the all models but the signal is formed faster in MBS. For the k values around 0.01 several pulses are produced in MBS and the first pulse is also faster relative to BS and HH.

### 9-3. Investigating the case of no external stimulus

Basically, we expect to have no spike generation without any stimulus. Now, we investigate this point when there is no stimulus. In the diagrams of Figure 4, the spikes have been initiated by a short current pulse (i.e. 1ms duration) acting at t <0.

If the amplitude of the stimulating current pulse is lower than some critical value, the membrane potential returns to the rest value without a large spike-like outing. The results for no stimulus, S = 0, are plotted in Figure 5. The diagrams show that the weak entanglement again creates additional spikes in MBS which is not realistic. Figure 5 shows that when there is no stimulus the amplitude of voltage in the HH and BS models are so small that we see them almost as a straight line like the rest state, but the amplitude for the MBS model is still like the state with stimuli. It indicates that the pulse in the MBS model is not stimuli dependent.

### 9-4. Investigating different lifetime stimuli

Now, we insert different lifetime stimuli to investigate the variations in the number of spikes. Here, we change the stimulating duration (SD) of external currents. Figure 4 was plotted for SD = 0.001ms. We plot other SD values as 0.003, 0.007, 0.009, 0.02, 0.05, 0.07ms for different k values as k = 0.01, k = 0.1 and k = 1 in Figure 6. The number of spikes are seen to grow when SD for k = 0.01 and k = 0.1. In this case, the difference between the models is that the BS and HH models overlap while the MBS is different. There is a small deviation between BS and HH models for the case k = 1.

As a result, in Figures 4, 5 and 6 there is a high overlap between BS and HH models whereas MBS is highly affected under different conditions. There is only a small deviation in few cases between BS and HH models in the hyperpolarization steps, which should be interpreted.

### 9-5. TEMPERATURE-DEPENDENT EQUATIONS

The original HH model applies a temperature of $6.3°C$. For higher temperatures, the gating variables include a temperature coefficient, X, so the activation (m, n) and inactivation (h) gating variables are given by the following differential equations (Guttman, 1972).

$$\frac{dY}{dt} = [\alpha_Y(V)(1-Y) - \beta_Y(V)Y]X \qquad (6)$$

where Y = m; n; h with the values already mentioned in the method. The temperature coefficient, X, for temperature $T[°C]$ is ( Guttman, 1972):

$$X = 3^{\frac{(T-6.3)}{10}}, T\ [\ °C] \qquad (7)$$

Now, we plot the voltage diagrams of the three models for temperatures $T = 6.3°C$, $T = 18°C$, $T = 20°C$ and $T = 25°C$ for k = 0.01, k = 0.1 and k = 1 in the Figure 7. It is seen that the spikes initiation becomes faster by increasing the temperature in the all models. For the case k = 0.01 the number of spikes dramatically increases in the MBS model while for k = 0.1 and k = 1 there is no any change in the number of spikes. For the higher degrees of entanglement, k = 1, the models become more similar. As a result, the consistency between BS and HH is high but MBS deviates largely.

### 10. DISCUSSION AND CONCLUSIONS

In this paper, two aspects relevant for action potential generation are discussed. The first aspect discussed the relation of filter gating to voltage sensitive pore gating, and the second is the replacement of classical probabilities (CPs) with quantum probabilities (QPs) in the Hodgkin-Huxley (HH) equation.

Our results indicate that the BS model has no significant discrepancy from the HH model under different conditions and in some cases, even has a better explanation, for example quantum effects may be the cause of quicker initiation of spikes in cortical neurons according to the paper published by Naundorf et al. (2006), despite the debate about this property (Mc Cormick et al., 2007). It has also been shown that neuromuscular junction transmission is affected by quantum effects of ion transition states according to the BS model (Rahman & Mahmud, 2012). However, despite the excellent results obtained by the BS model it fails to account for a reasonable relation between the QPs and the voltage of the channel. Moreover, these QPs do not explain the configuration of ions in the SFG. If we put the real configuration of ions in the SFG, the BS model becomes the MBS model. The comparisons show that the MBS model will be dramatically different (see Table 1) and unrealistic. In fact, MBS is only an indicator model, which is based on the configuration of ions in the selectivity filter, for comparison with the BS and HH models. Our results indicate that the MBS model cannot be a good model; it deviates from the HH model significantly and dies not explain the spike generation correctly – and consequently the quantum probabilities cannot be expressed in terms of ion locations in the selectivity filter. On the other hand, the BS model is expressed based on the degrees of freedom of activation states and this expression is a good chance since not only dies the BS model not deviate from the HH model under different conditions, it can even provide better results relative to the HH model in some cases.

Indeed, the BS model introduces a semi-quantum-classical version of the classical HH equation for propagating neuronal voltage pulses (action potentials, spikes), which predictably and reproducibly changes the onset characteristic of the signal in line with the direction that is found in real cortical neurons. The BS version of the HH model is strongly suggestive for cooperative, quantum chemical events that combine ion selectivity, permeation rates and classical gating states. However, neither the classical HH model nor the BS version explicitly claim or address the physical identification of these events. In particular, there must be a yet unknown mechanism that combines selectivity filter states of single atoms with voltage dependent pore gating states of the protein. This kind of interaction has to involve a severe quantum-classical transition, crossing many action orders within the protein and its atomic environment.

We should note here that the role of SFG in the channel is serious and it is in principle possible to integrate SFG properties in the HH equation. However, there is still no such a comprehensive model to explain both classical and quantum mechanical aspects. Consequently, to develop a mathematical equation including quantum effects to explain the pulse generation we should take account of noise, structure and configurations of ions in the SFG as well as an explanation for a relation between the QPs and the voltage. Our results show that the semi-quantum-classical BS model is more compatible with experiments although it does not explicitly provide a physical description of ion configurations. This could be a prospective subject for future investigations in this context. Physical modelling along this direction may, however, gradually help to identify the nature of the underlying gating weights and help to disclose a possible quantum nature at the microscopic scale that propagates into the classical HH equations of motion, as suggested in the BS version.

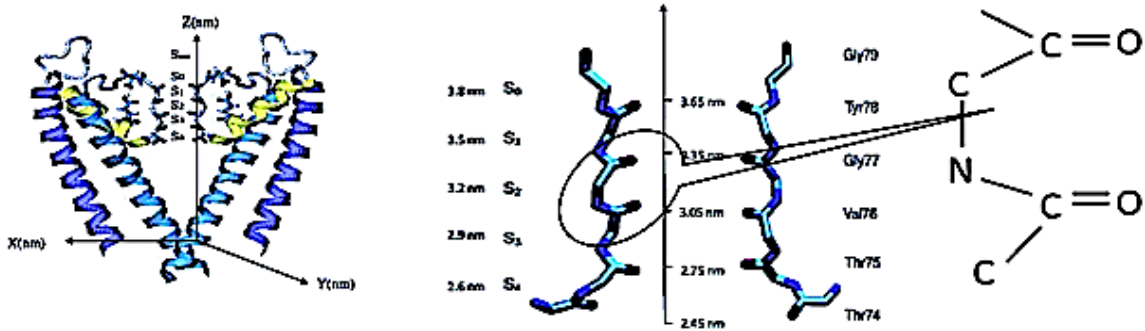

FIG. 1: Left side) A representation of KcsA ion channel. Center) Two KcsA-P-loop monomer chain in the selectivity filter, composed of the sequences of TVGYG amino acids [T(Threonine, Thr75), V(Valine, Val76), G(Glycine, Gly77), Y(Tyrosine, Tyr78), G(Glycine, Gly79)] linked by peptide units H-N-C=O. Right side) Representation of carbonyl (C=O) groups in the selectivity filter.

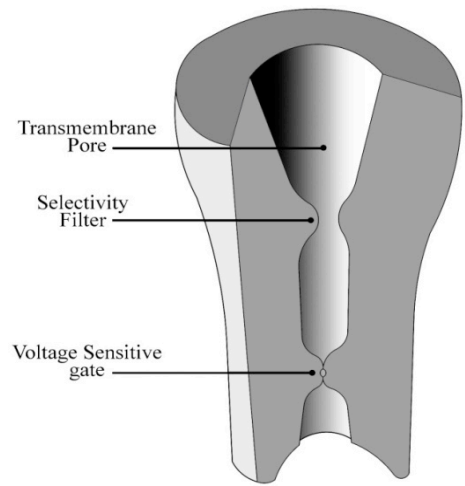

FIG. 2: Representation of the selectivity filter gate (SFG) and voltage sensitive gate (VSG) in a simple scheme of the ion channel.

| Model | NS T=6.3[°C] t=0.05(s) | NS T=18[°C] t=0.05(s) | NS T=20[°C] t=0.05(s) | NS T=25[°C] t=0.05(s) | NS′ SD=0(s) t=0.03(s) | NS′ SD=0.003s t=0.03(s) | NS′ SD=0.007(s) t=0.03(s) | NS′ SD=0.009(s) t=0.03(s) | NS′ SD=0.02(s) t=0.03(s) | NS′ SD=0.05(s) t=0.03(s) |
|---|---|---|---|---|---|---|---|---|---|---|
| HH | 1 | 1 | 1 | 1 | $\cong 0$ | 1 | $\geq 1$ | $\geq 1_*$ | $\geq 2$ | 3 |
| BS k=0.01 | 1 | 1 | 1 | 1 | $\cong 0$ | 1 | $\geq 1$ | $\geq 1_*$ | $\geq 2$ | 3 |
| BS k=0.1 | 1 | 1 | 1 | 1 | $\cong 0$ | 1 | $\geq 1$ | $\geq 1_*$ | $\geq 2$ | 3 |
| BS k=1 | 1 | 1 | 1 | 1 | $\cong 0$ | $1\ast$ | $\geq 1\ast$ | $\geq 1\ast_*$ | $\geq 2\ast$ | $3\ast$ |
| MBS k=0.01 | 5.5 | 17.5 | $\geq 19$ | 34.5 | 3.5 | 3.5 | $\geq 3.5$ | $\geq 3.5_*$ | 4 | $\geq 4$ |
| MBS k=0.1 | 1 | 1 | 1 | $\geq 1$ | 1 | 1 | 1 | 1 | 1 | 1 |
| MBS k=1 | 1 | 1 | 1 | 1 | 1 | 1 | 1 | 1 | 1 | 1 |

TABLE I: Comparison of HH, BS and MBS models in number of spikes (NS) under different conditions. The symbol $*$ indicates that $NS \geq 1$ the but more than when we have SD=0.007(s) and the symbol $\ast$ shows the variation in the form of spikes with creation of smaller spikes. The symbol "$\geq 1$" means a value between 1 and 2, thus the symbol "$\geq$" indicates between two integer (or half integer) number. In the all models, we have: $T_{HH}, T_{BS} \geq T_{MBS}$ and $f_{HH}, f_{BS} \leq f_{MBS}$ for firing frequency, and the width of the major spikes decrease by increasing temperature. The symbol ′ means that the temperature T=6.3°C.

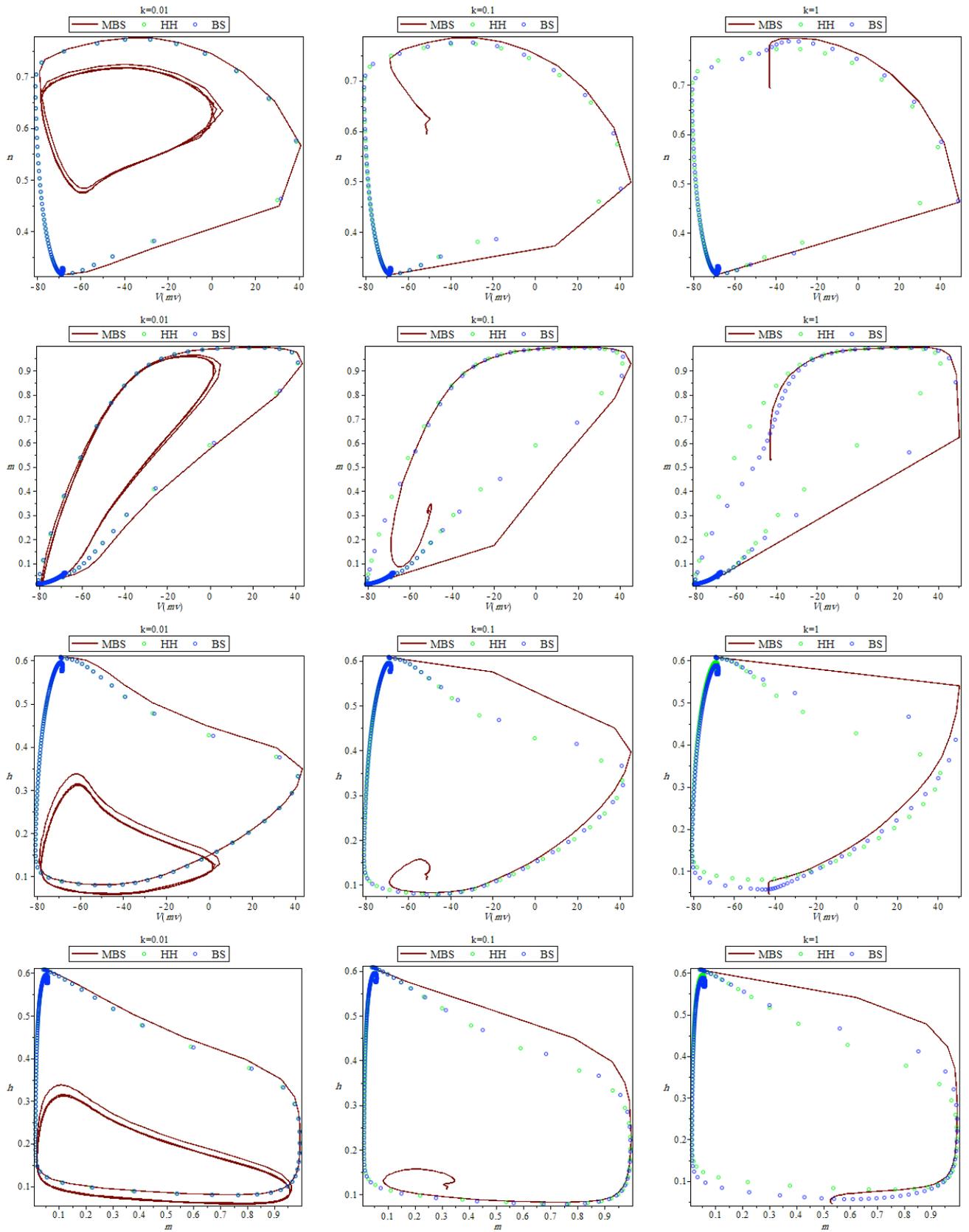

FIG. 3: Dynamics of gate variables m, n and h for the HH, BS and MBS models for different k values.

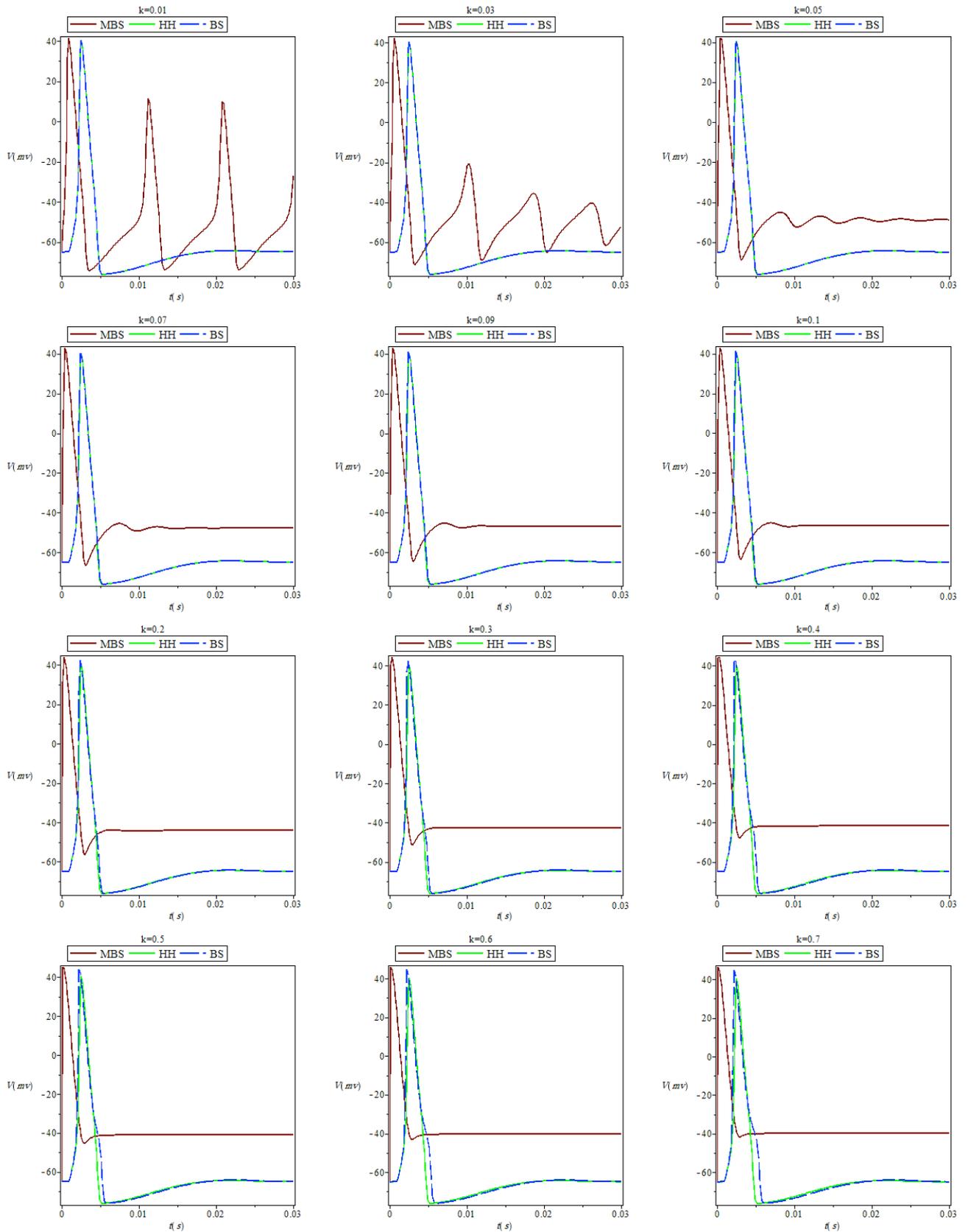

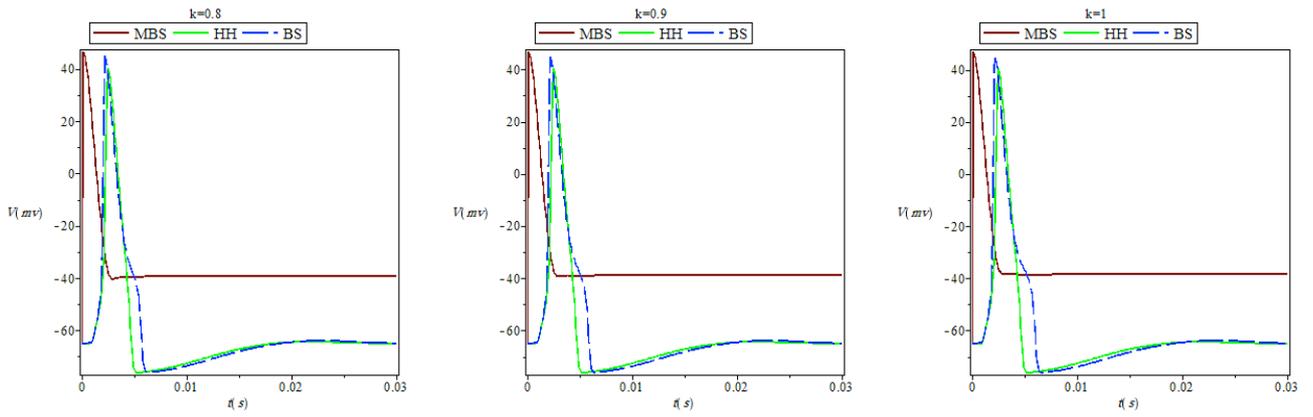

FIG. 4: The voltage diagrams versus time for different values of k.

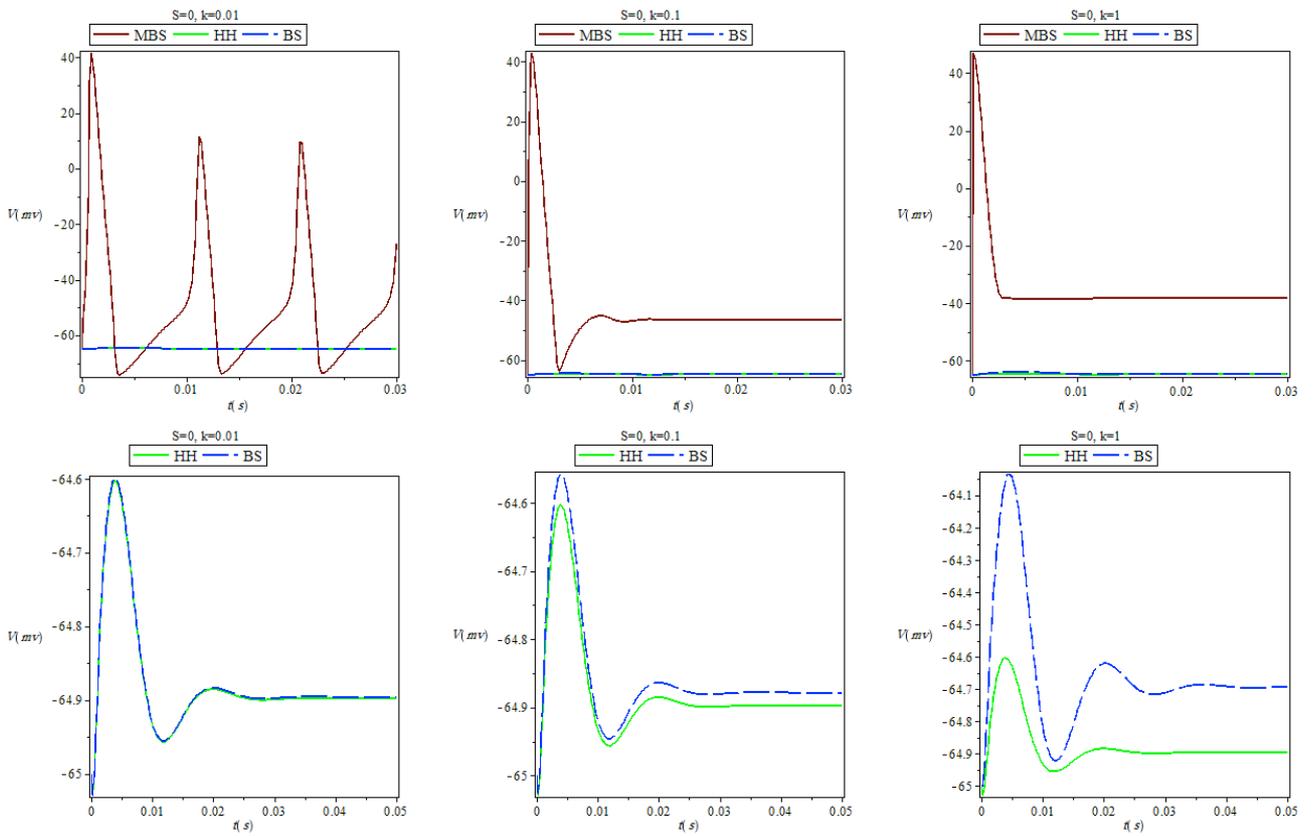

FIG. 5: The voltage diagrams versus time when there is no stimulus for different values of k.

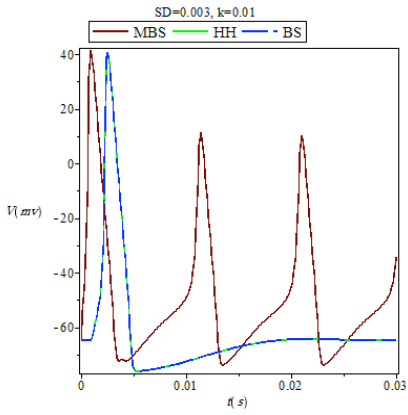
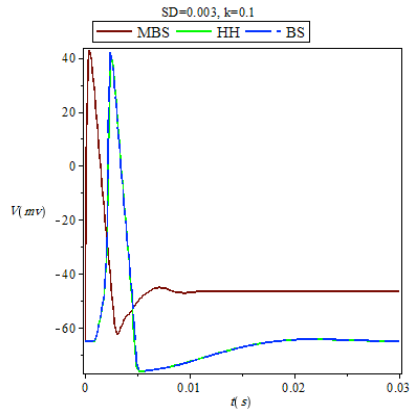
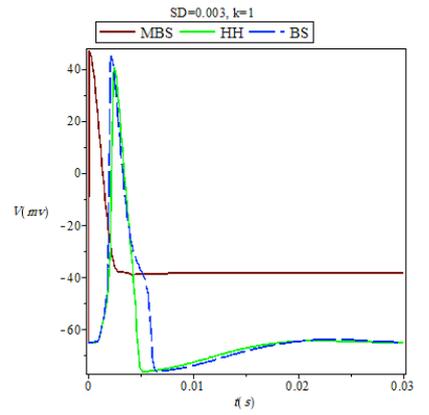
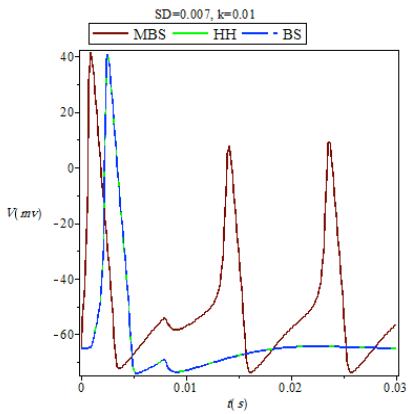
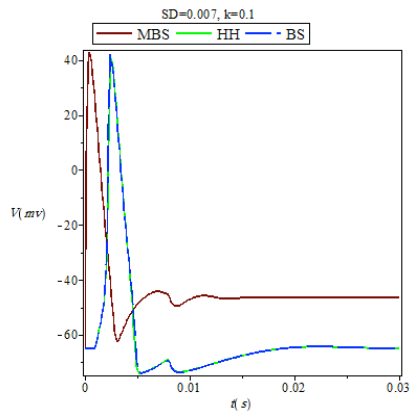
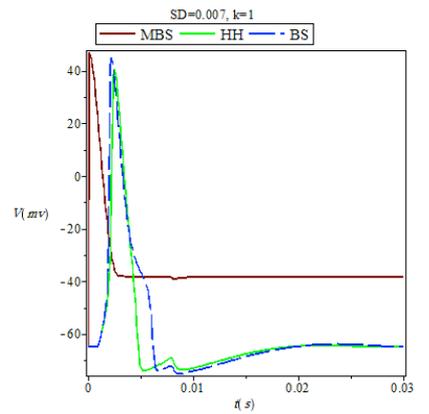
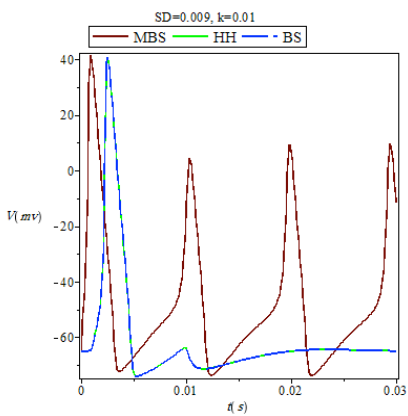
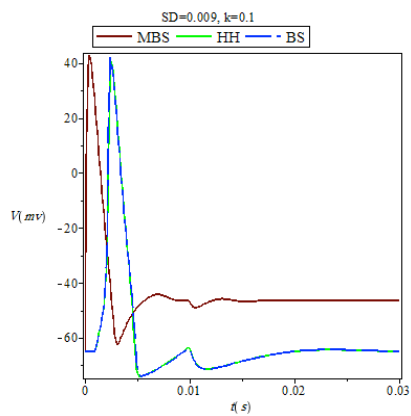
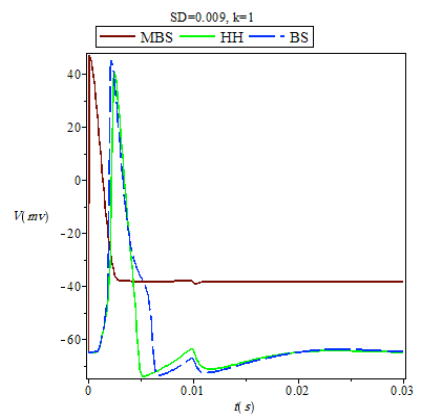
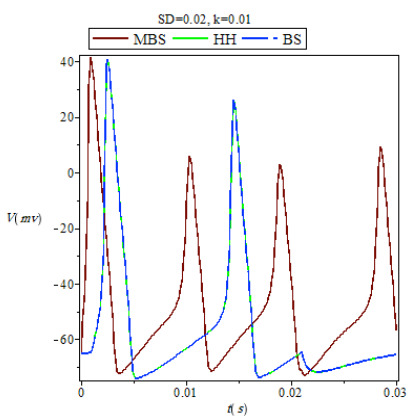
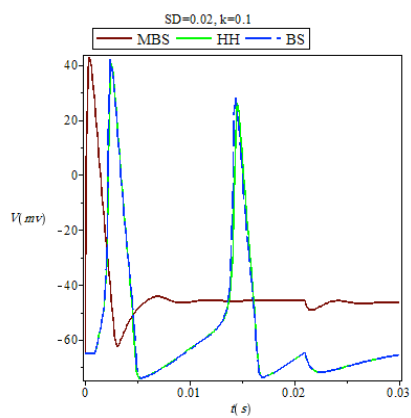
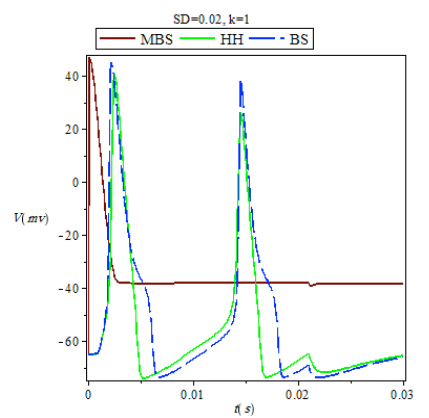

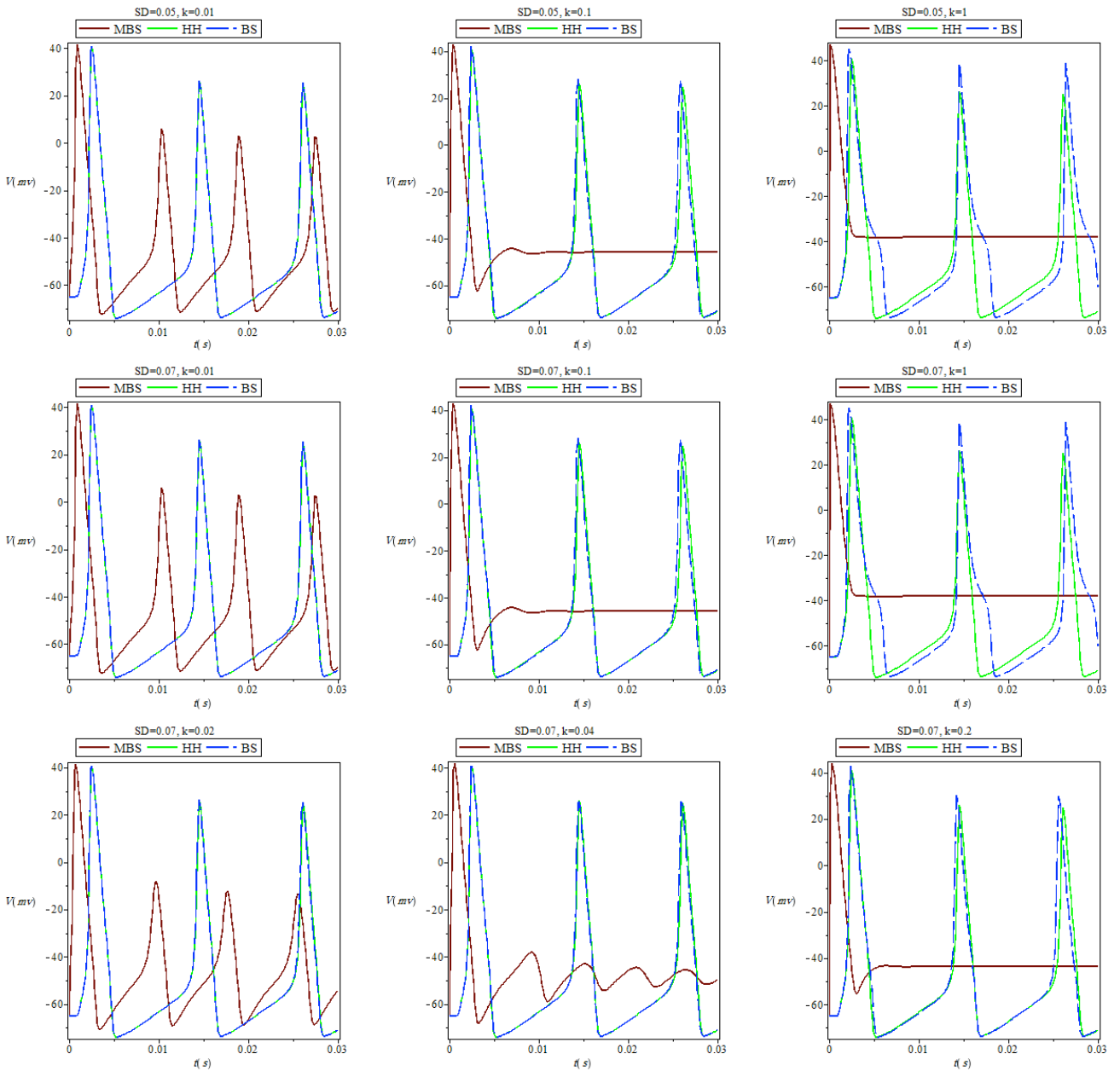

FIG. 6: Voltage diagrams for different stimulating current pulse duration(SD) for different values of k.

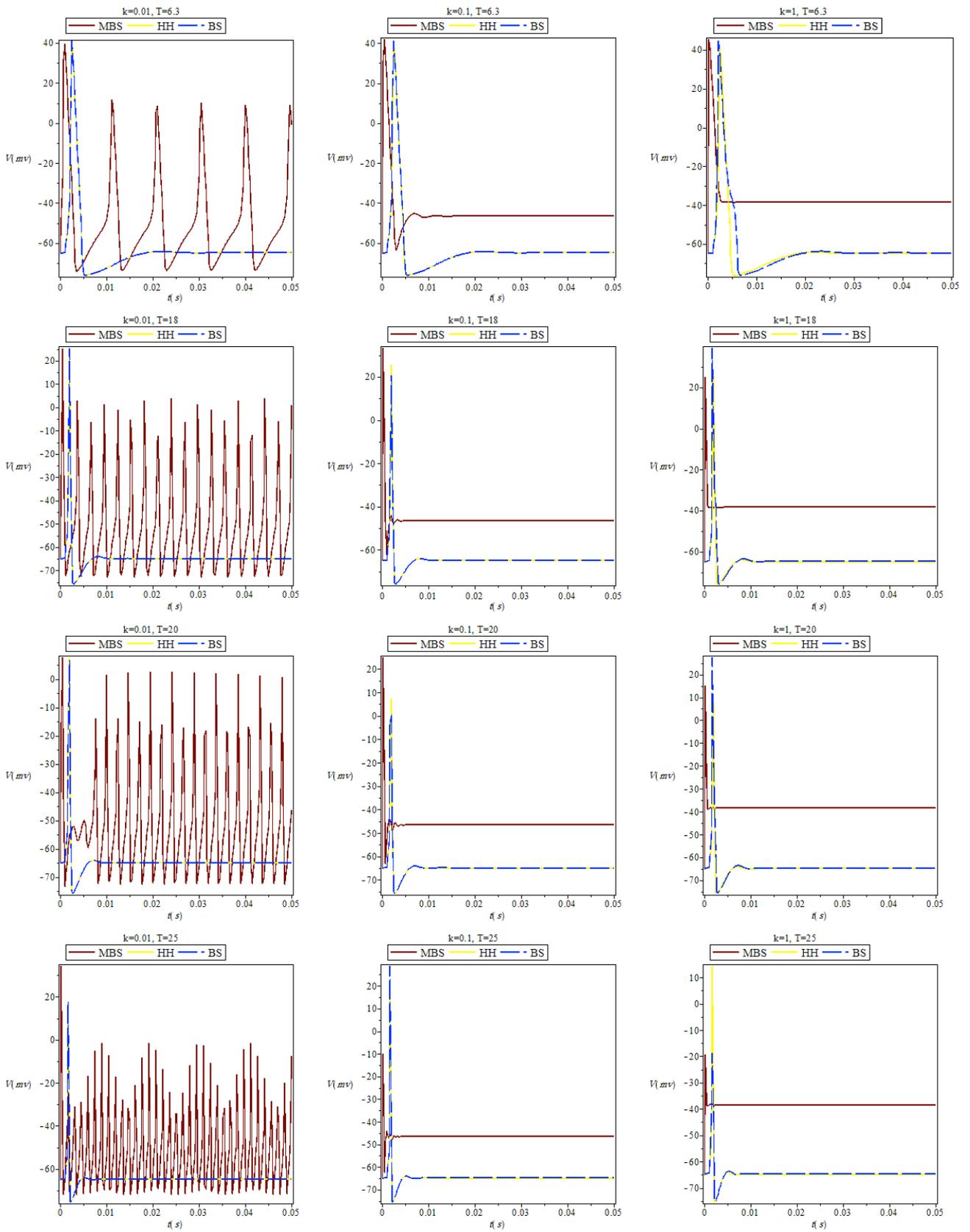

FIG. 7: Voltage diagrams at different temperatures for different values of k.